\documentclass{aa} 
\usepackage{graphicx}
\usepackage{natbib}
\def\simgt{\lower.5ex\hbox{$\; \buildrel > \over \sim \;$}}
\def\simlt{\lower.5ex\hbox{$\; \buildrel < \over \sim \;$}}
\newcommand\gad{$\nabla_{\rm ad}$}
\newcommand\teff{T$_{\rm eff}$}

\newcommand\msun{M$_{\odot}$}
\newcommand\ocen{$\omega$~Cen~}

\begin{document}
   \title{A detailed study of the  main sequence of the Globular Cluster NGC~6397: \\
  \subtitle{ can we derive constraints on the existence of multiple population?}}

   \author{Di Criscienzo, M.,\inst{1},
          D'Antona, F.\inst{1}
          \and           
          Ventura, P.\inst{1}
          }

   \institute{\inst{1} Osservatorio Astronomico di Roma,
              Via di Frascati 33, 00040, Monte Porzio Catone, Rome, Italy\\
              \email{dicrisci, dantona, ventura@oa-roma.inaf.it}}

   \date{accepted 15/12/2009}

\abstract
{}
{If NGC~6397 contains a large fraction of ``second generation" stars ($>$70\% according to recent analysis), 
the helium abundance of its stars might also be affected, show some star-to-star variation, and
be larger than the standard Big Bang abundance Y$\sim$0.24. Can we derive constraints on this issue
from the analysis of the main sequence width and from its luminosity function?}
{We build up new models for the turnoff masses and the main sequence down to the hydrogen
burning minimum mass, adopting two versions of an updated equation of state (EOS)
including the OPAL EOS. Models consider different initial helium and CNO abundances to cover
the range of possible variations between the first and second generation stars. We compare the
models with the observational  main sequence. We also make simulations of
the theoretical luminosity functions, for different choices of the mass function
and of the mixture of first and second generation stars, 
and compare them with the observed luminosity function,
by means of the Kolmogorov Smirnov --KS-- test. }
{The study of the widht of the main sequence at different interval of  magnitude is consistent with the hypothesis that both generations are present in the cluster. If the CNO increase suggested by spectroscopic observation is taken into account  the small helium spread of the main sequence in NGC~6397 implies a substantial helium uniformity ($\Delta$Y$\sim$0.02) 
between first and second generation stars. The possible spread in helium doubles if an even larger increase of CNO is considered.
The luminosity function is in any case well consistent with the observed data.} 
{}
\titlerunning{The helium content in the Globular Cluster of NGC~6397}
\authorrunning{Di Criscienzo et al.}
\maketitle

%

\section{Introduction} 

Our views about Globular Clusters (GCs) are dramatically changing in these latter years, 
thanks to precise photometric investigations, that revealed the presence of multiple 
main sequences or subgiant branches \citep[e.g.][]{piotto2007, milone2008}, and the 
increasing amount of new spectroscopic data on GC stars. These data, in particular 
the spectra for about 2000 stars in 19 GCs recently obtained
by the multiobject spectrograph FLAMES@VLT \citep{carretta2009a,carretta2009b}
have shown how the ``chemical anomalies" among GC stars are indeed ubiquitous in all clusters, 
and concern a large fraction (from 50 to 80\%) of stars. The peculiar chemical abundances take the form of anticorrelations between O 
and Na, Al and Mg, and are present in stars of various evolutionary phases (both 
unevolved main sequence and evolved red giant branch stars) \citep{gratton2001, 
ramirez2002, carretta2004}, supporting the idea that they are not due to 
an ``in situ" deep mixing in the stars, but have been imprinted in the 
gas from which they formed, that was polluted by the winds lost by a previous 
first generation (FG) of stars: according to this hypothesis, we are
now seeing a second generation (SG) of stars mixed to the FG. 
It is not definitely settled 
what kind of stars produced this material, that must have been processed by the 
hot CNO cycle and other proton capture reactions on light nuclei. The two most 
popular candidates are massive asymptotic giant branch (AGB) stars \citep[e.g.][]{ventura2001},
and possibly, for the extreme anomalies, super--AGBs  \citep{pumo2008},
or massive stars, either fastly rotating \citep[e.g.][]{decressin2007}, or in binaries 
undergoing non conservative evolution \citep{demink2009}.\\
\cite{carretta2009a} examine the famous Na--O anticorrelation  and 
tentatively divide the stars of each cluster in three groups: the ``primordial" stars, those
having abundances of O and Na similar to those found in the halo stars; the ``intermediate"
stars, having high Na and somewhat depleted O; the ``extreme" stars, having high 
Na and strongly depleted O. While only a few, very massive, clusters contain stars  with 
extreme anomalies, all clusters show a population with ``intermediate" chemistry.
The problem of GC formation and early evolution is very complex, but it is easy to accept
that a cluster contains two or multiple populations, if it shows both chemical peculiarities, as
discussed above, and photometric peculiarities: the GC NGC~2808 is a prototype of this 
class\footnote{We do not wish to include \ocen among the classic GCs, as it also shows large
metallicity variations, indicating that its evolution is partially similar to that of a massive GC,
containing, e.g., a population with a very high helium content, but also closer to that of a
small galaxy, as the supernova ejecta take part in the star formation events.}, as it shows three 
populations both in the main sequence \citep{piotto2007} and in its horizontal branch (HB) 
extended morphology \citep{dc2004}, and it is possible to reproduce these 
three populations by assuming that they differ in helium content \citep[e.g.][]{dc2008}; 
in addition, this cluster shows the Na--O anticorrelation in one of its most
extreme forms, with stars reaching very low oxygen abundances. \citet{carretta2006} even find a possible
indirect hint of helium enhancement in the subgroup of oxygen--poor red giants that they examine.
As theory expects that some helium enrichment accompanies the hot--CNO nucleosynthesis in both the
proposed models (massive AGBs or massive stars), NGC~2808 represents in many respects, the most classic example of
well understood multiple populations in GCs.

There are other cases, however, in which the presence of subpopulations is not as well clear, 
and, in particular, it is not clear whether the abundance anomalies \citep[in some cases less prominent,
but always present, according to][]{carretta2009a} are {\it always} accompanied by helium enrichment,
and how large this enrichment is. In particular, the cluster NGC~6397 has always been considered the
perfect example of a ``simple stellar population" (SSP) due to the ``tightness" of its HR diagram, including
a very compact blue HB. In recent years, spectacular 
data for the low main sequence of NGC~6397 have become available. By using the 
technique of proper motion cleaning, \cite{richer2006} obtained a very tight main 
sequence and a very clean luminosity function down to the hydrogen burning minimum
mass (HBMM) \citep{richer2008}. We decided to use these data to quantify 
at what level we can accomodate helium variations in this cluster (and the possible,
associated, CNO variations), by investigating whether the main sequence width and its
luminosity function are compatible with a helium spread and/or with a non--standard helium content.
At the same time, we take the occasion of this comparison to 
compute and test new stellar models for the low main sequence. In these models,
we employ and compare new equations of state (EOS) today available, and we 
test the available color-T$_{\rm eff}$ transformations.\\
The outline of the paper is the following. In Section
2 we describe in some detail the spectroscopic results concerning the cluster, and what we expect
concerning the multiple populations it should hide, and what are plausible helium and C+N+O variations 
expected. We then summarize the photometric data by \cite{richer2008}. After having summarized the literature concerning the
low mass main sequence models, in Section 3 
and 4 we describe our  code and the results of the computation
of solar scaled models; in Section 5 we present $\alpha$-enhanced models computed for the comparison  with the data for the Globular Cluster NGC~6397. 
Both the comparison of the CMD and the luminosity function with theory are discussed in 
detail in Section 6 also in the hypothesis of multiple populations with different helium, 
and possibly also C+N+O, content. In Section 7 we summarize our results and conclusions.


\section{The case of NGC~6397: apparently, a simple stellar population}
\cite{carretta2009a} find that at least 70\% of stars in the cluster 
NGC~6397 are ``intermediate" according to their definition, 
although photometric studies show that all the 
evolutionary sequences in this cluster look like those of a prototype SSP:
the main sequence is very tight 
\citep{king1998,richer2006}, and the horizontal branch (HB) 
lacks the extreme HB and blue hook stars that are now 
regarded as the proof of the presence of a very helium enriched population 
\citep{dantona2002,dc2004}. The chemical anomalies of \cite{carretta2009a} analysis, however, 
do not come as a full surprise, as already many hints were available in the 
recent literature about the dubious simplicity of this cluster. First of all, 
already \cite{bonifacio2002} had noticed the presence of Nitrogen rich stars 
that have a normal Lithium content \citep[see also][]{pasquini2008}, and 
\cite{carretta2005} find that only three subgiants out of 14 stars are nitrogen 
normal. These features lead us to suspect that the material from which these 
stars formed is CNO processed, as expected in the stars with low oxygen and 
high sodium. The possible helium enhancement is certainly not extreme, as, for instance, 
are small  lithium variation among the turnoff stars \citep{lind2009,korn2007,pasquini2008}.
In the massive AGB model for the formation of the second generation, the lithium content 
of the AGB ejecta is not extremely depleted as in the other models, but it is difficult
to believe in a cosmic conspiracy producing exactly the same lithium of the FG, unless the AGB matter
is very diluted with FG gas, so that both lithium and helium do not differ too much in the two 
generations. \\
A different, mostly theoretical, approach, led \cite{caloi-dantona2005} and 
then \cite{dc2008} to provocatively propose that {\it all} the stars in the 
clusters having entirely blue HBs are composed by SG stars. This idea is at the 
basis of a possible explanation of the peculiar difference between the GCs M~3 
and M~13. The cluster M~3 has a complex HB, including many stars redder than 
the RR Lyrae (a red clump), RR Lyrae stars, and a well populated blue side, while M~13, 
having the same metallicity, has an only--blue HB. This difference, that has 
been generally attributed to different age \citep{jb1998,rey2001} or to 
different mass loss along the red giant branch 
\citep{lee1994,catelan1998} (the famous second parameter problem) 
may also be interpreted by assuming that M~13 is totally deprived of its FG, and the 
SG has a minimum helium abundance in mass fraction Y$\sim$0.28. In M~13, whose HB 
shows a prominent blue tail, simulations of the HB stellar distribution show 
that there must also be a small fraction of stars with helium Y$\sim$0.38 
\citep{dc2008}. 
Is it possible that a GC is composed {\it only} of SG stars? This could indeed 
happen, as shown in some hydrodynamic plus N--body simulations of the SG 
formation and of the cluster first phases of dynamical evolution presented by 
\cite{dercole2008}. They find that the dynamical evolution of the cluster may be
characterized by expansion of the FG star system due to the SNII mass loss 
preferentially occurring in the cluster central regions, while the SG is still 
forming in the core. Depending on the initial conditions, in some cases, the 
ratio of SG to FG stars may even reach a factor 6 or more. While this kind of
modelling depends on the input parameters, and  
does not imply that this really occurred in nature, NGC~6397, having a 
short blue HB, and a tight main sequence (MS) and red giant branch (RGB), 
could indeed be made by a homogeneous set of SG stars, corresponding to a unique 
value of Y, just a bitter larger than the Big Bang abundance (Y$\sim$0.26-0.28). This 
speculation would help to understand why the nitrogen abundance in most of the NGC~6397 
stars is quite large. 
The possibility that all the stars in this cluster have a homogeneous, but larger than standard, 
helium abundance could be falsified by looking at the main sequence width, that depends on the
combination of the photometric errors with the possible star to star differences in helium.\\  
Apart from this extreme and provocative suggestion, very recently
\citet{carretta2009a} show that only up to $\sim$30\% of stars should belong to 
the primordial population (the FG), in  the assumption that all the stars within 
three sigma from the lowest sodium abundance measured in a cluster are ``primordial''\footnote{
Indeed, while  \citet{carretta2009a} have only 4 ``primordial'' stars in their sample 
of O-Na measurements for this cluster, they have determined Na abundances in a larger number of stars, 
and also in this larger sample sodium is ``normal'', that is similar to the sodium of
the halo stars having similar metallicity, in $\sim$ 25-30$\%$ of stars.}. So, at present, 
the most reasonable assumption is that NGC~6397 has at least 70$\%$ of SG.\\
Given the sodium and oxygen abundances of the anomalous cluster stars, do we expect that they 
have an enhanced helium content? In the hypothesis  that the SG is born from matter 
mixed with the hot--CNO processed ejecta of massive AGBs, we can look at the results by 
\citet{vendan2009}. They interpret the anomalous Na--O abundances in NGC~6397 as a result 
of mixing between  50\% of pristine gas with 50\% of gas ejected by the 5M$_{\odot}$ AGBs. 
Looking at Table 2 of their paper, the helium abundance in the 5M$_{\odot}$ ejecta for Z=0.0006 
is Y=0.329. A dilution by 50\% with matter having primordial Y=0.24 provides indeed Y=0.285.
If we take these results at face value, also the total CNO content of the SG stars is larger than that
of the FG stars. The 5\msun\ AGB evolution provides a CNO enhancement by a factor $\sim$3, so  
we must consider also increased CNO (and total metallicity) by a factor 1.5, when we compute the
higher helium models.  
Of course, the computed AGB models do not give a mandatory prescription of what really happens in the cluster, so that we will also consider
normal CNO models and models with even larger CNO in order to include all possible cases.
Helium abundances equal or larger than the standard one will be considered, up to Y=0.28,
to understand whether the hypothesis that at least $\sim 70$\% of stars 
in NGC~6397 have an helium abundance larger than the Big Bang abundance
is consistent with the photometric data.

\subsection{The observational color-magnitude diagram of Globular Cluster NGC~6397}
The first HST observations of the low MS of NGC~6397 date back to \citet{paresce1995}. 
Afterwards, \citet{king1998} have observed  NGC~6397 with WFPC2@HST finding that the 
luminosity function has a rapid decline at low mass end.
Recently, the deepest observations by \citet{richer2006,richer2008} refined the data. 
They observed an outer region of NGC~6397 with 
ACS@HST using the photometric filters F814W and F606W. The high sensitivity of the 
camera, the large number of orbits obtained (=126) and the vicinity of the 
cluster (it is the second closest globular cluster, after M4)  made it possible 
to reach the deepest intrinsic luminosities for a globular cluster achieved until today, and what appears as the termination of the MS.
The field 
observed overlaps that of archival WFPC2 data from 1994 and 1997, which were 
used for the  proper-motion-cleaning of the data. This technique, applied to the deep 
ACS photometry, produces a very narrow MS till its end.
These observations are a good basis  to test the physics 
of low mass stars and the possible role of the helium
abundance. 
\citet{richer2008} analyzed the color-magnitude diagram (CMD) diagram using the results  
of models computed with Dartmouth Stellar Evolution Program 
(DSEP) \citep{dotter2007}. Their main results are that:
\begin{itemize}
\item the MS appears to terminate close to the CMD location of the HBMM 
predicted  by models. The authors state that they 
would have found fainter MS stars  in the cluster, if there had been any;

\item the MS fitting technique provides a good agreement  down to 
F814W=22.5 mag; below this, down to $\sim$24 mag, the isochrone is either too 
blue or too low in luminosity. The authors underline that this is likely due to low mass 
models being less luminous at a given mass than real stars;

\item exploring the MS luminosity function, they find  that a power law for the mass 
function MF well reproduces the distribution in luminosity of the observed stars, 
while a more top--heavy MF is necessary to fit the data in the cluster core  that they have 
also available. However, theory predicts more stars than 
observed at the lowest MS luminosities, as also previously found by \citet{montalban2000}
using data by \citet{king1998} 
\end{itemize}

\section{The low mass main sequence models:Input Physics of the models}
The computation of very low mass stellar models requires an
accurate knowledge of the EOS for partially ionized gas at 
densities where the ideal gas EOS approximations breaks down dramatically, 
especially close to the pressure ionization region 
\citep{fgvh1977,mm1979,saumon1995}. The general properties of very low masses 
have been described in many works \citep[for example see the reviews by]
[]{chabrier1997,alexanderetal,cassisi2000}; for the population II low masses, after the work by 
\cite{dantona1987}, based on grey atmosphere boundary conditions and on the EOS 
by \cite{mm1979}, \cite{baraffe1997} presented models based on the 
\cite{saumon1995} EOS and on the NextGen non-grey atmosphere models later on 
published by \cite{hauschildt1999}. They showed that this latter improvement was 
essential to reproduce the colors of the low mass main sequence. \cite{montalban2000}, reexamining the problem of the EOS, 
cautioned about the use of the Additive Volume interpolation needed to obtain 
the thermodynamic quantities for intermediate compositions from the pure 
hydrogen and pure helium tables available in the \cite{saumon1995} EOS. In 
recent years, two new EOS have become available: the FreeEOS by \cite{irwin2004} and the OPAL 
EOS, namely the EOS provided by the Livermore group as a byproduct of the 
opacity computation \citep{rogers1996}. Both EOS however do 
not cover the pressure ionization region, for which the best approach remains 
that by \cite{saumon1995}.The FreeEOS 
has been recently employed by \cite{dotter2007} 
and applied to the fit of the main sequence of NGC~6397. 
The OPAL EOS has not yet been used to approach the construction 
of low mass models, so we decided to adopt it in two different ways in our new models,
as described in Sect.~4.\\
We use the ATON code for stellar evolution; a detailed description can be found in 
\citet{ventura2007}; in the following we remember the main updated inputs
that are important for the treatment of low mass stars.

\subsection{Opacities and nuclear  reaction}
The program includes the opacities by \citet{ferguson2005} 
for the external region of the star (T$\le$15000~K)  and the latest version 
(2005) of OPAL opacities for higher  temperatures \citep{iglesias1996}. For fully convective   
low mass stars (below $\sim$0.35M$_{\odot}$), 
the  uncertainties on  radiative opacities have negligible influence on the models,
while for larger masses an error in opacity by $\sim$20$\%$
may cause an error up to $\sim$1$\%$ in the determination of  radii \citep{dotter}. 
Electron conduction opacities were  taken  from the WEB site of Potekhin (2006) and correspond to the \cite{potekhin1999} treatment, corrected following the improvement of the treatment of the e-e scattering 
contribution described in \citet{cassisi2007}.  \\ Although not necessary in this
computation, the nuclear network includes 
30 chemical elements, all main reactions of p-p, CNO, Ne-Na and Mg-Al chains 
and the $\alpha$ capture of all nuclei up ${^{26}}$Mg. The relevant cross section are 
from the NACRE compilation \citep{angulo1999}.\\

\subsection{Equation of state: OPAL {\it vs} \citet{saumon1995}}

\begin{figure}
\centering
\vspace{-20pt}
\includegraphics[width=8cm]{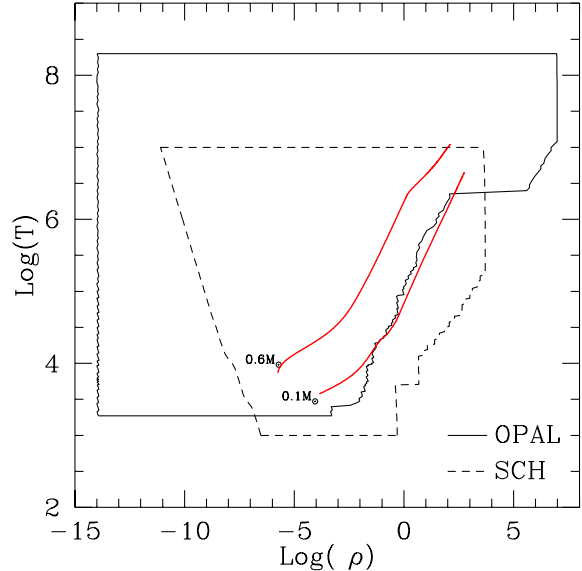} 
\vspace{-50pt}
\caption{We show the region of the  Log($\rho$)-Log(T) plane (density $\rho$ in g~cm$^{-3}$ and temperature T in K) where OPAL (solid box) 
and SCH (dashed box) EOS are available. The structure profile of two stars 
with Age=10 Gyr, Z=0.0006 and with M=0.6  M$_{\odot}$ and 0.1 M$_{\odot}$ 
are also shown. The 0.6M$_{\odot}$ is in the region of overlapping of the two EOS, while 0.1M$_{\odot}$ is exactly at the border between the two EOS.} 
\label{figEOS}
\end{figure}
\begin{figure}
\centering
\vspace{-20pt}
\includegraphics[width=8.5cm]{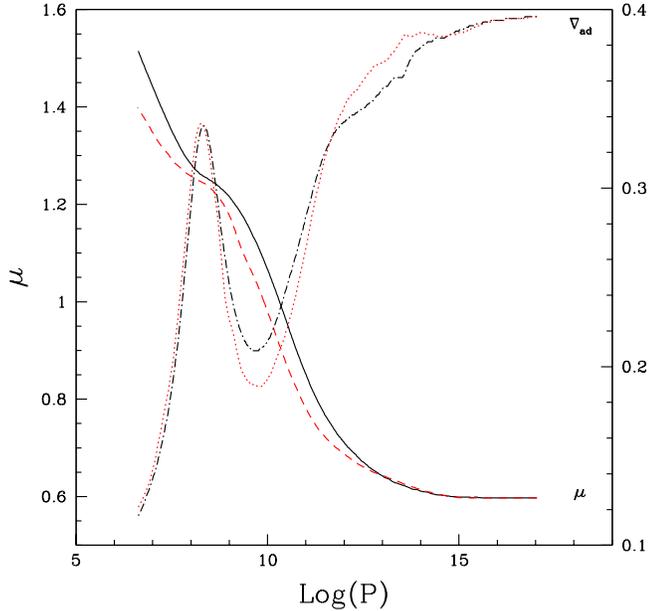} 
\vspace{-50pt}
\caption{Adiabatic gradient (\gad; dash-dotted and dotted curves; right {\it y}-axis) and molecular weight ($\mu$, solid and dashed curves; left {\it y}-axis) as a function of pressure P (cgs units) for the structure with M=0.3 M$_{\odot}$, Z=0.0006, Y=0.24 and Age=10 Gyr. The solid and dash dotted curves  correspond to EOS+OPAL while dashed and dotted curves are calculated with EOS+SCH (see text).} 
\label{figEOS2}
\end{figure}
\begin{figure}
\centering
\vspace{-20pt}
\includegraphics[angle=0,scale=.4]{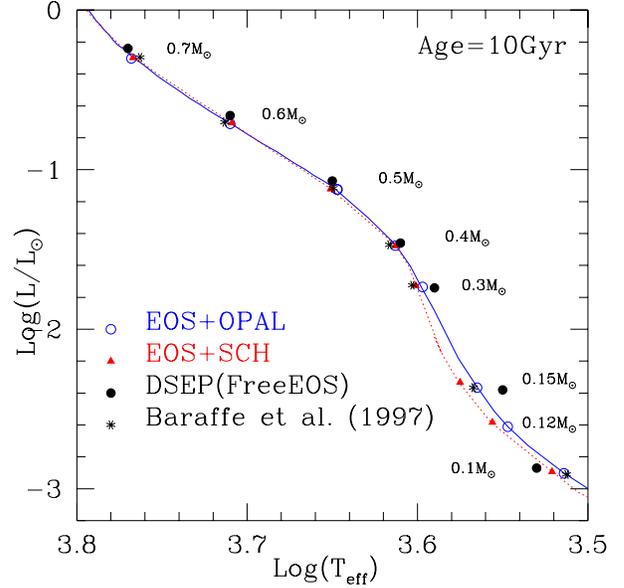} 
\vspace{-50pt}
\caption{HR diagram location at 10 Gyr for models 
with Z=0.0006, the labelled masses and calculated with different EOS tables. 
Models from  the Dartmouth 
Stellar Evolution Database (full dots) and from Baraffe et al. (1997) (asterisks), with same metallicity of ours, are also shown for comparison. } 
\label{figEOS3}
\end{figure}

 
As remarked previously, non-ideal effects become increasingly important for 
masses M$\simlt$ 0.8M$_{\odot}$. \\ 
ATON uses 18 tables of EOS in the (gas)pressure-temperature plane corresponding to three 
different metallicities, Z=0, 0.02 and 0.04, and six hydrogen mass fractions X, ranging from 
0 to 1-Z; the thermodynamic quantities, i.e. density, adiabatic gradient, specific heat at 
constant pressure C$_p$, the C$_p$/C$_v$ ratio, 
and the three exponents $\Gamma_1$, $\Gamma_2$ and $\Gamma_3$ are obtained via 4 cubic 
unidimensional splines on X, Z, pressure and temperature.\\ 
These 18 tables are built up in three steps. First, the thermodynamic quantities  
are computed according to the formulation by \citet{stoltzmann2000}, which is the most modern 
and updated description available for ionized gas, including both classic and relativistic 
degeneracy, coulombian effects and exchange interaction. The tables are then 
partially overwritten  by the OPAL EOS in the whole domain where this is 
available \citep[see OPAL WEB page, last update in February 2006]{rogers1996}.
Finally in the very low-temperature regime, where OPAL EOS is not available 
(see Fig. \ref{figEOS}) the tables are overwritten by the \citet{saumon1995} EOS, 
which has the advantage of employing an adequate physical model for the pressure 
ionization. The Saumon et al. EOS is only given for pure hydrogen and 
pure helium mixtures; the presence of metals is thus simulated by adding helium, 
and the different H-He mixtures must be interpolated through 
the Additive Volume law.\\ 
We call ``EOS+OPAL" these tables, that represent the standard in our computations.
To investigate how the results depend on the chosen EOS, we 
built additional tables (EOS+SCH) using the EOS by \citet[SCH]{saumon1995} 
in the whole region of the P-$\rho$ plane for which it is available. This is an 
interesting test, because the structure of the majority of stars discussed here 
are contained in the region of plane LogT-Log$\rho$ where both EOS are available 
(see Fig. \ref{figEOS}). \\ 
In Fig. \ref{figEOS2} we show the differences in the molecular weight ($\mu$) and 
adiabatic gradient \gad=d$\log$P/d$\log$T along the structure of two main sequence models 
of M=0.3M$_{\odot}$, Z=0.0006 at the age of 10Gyr, computed with the two different 
EOS. The differences are more evident in the zone of partially ionization where 
physical differences of the two treatments (different $\mu$) affect \gad. In the 
EOS+SCH models, the regions in which \gad\ is lower prevail, and the global effect is to 
produce \teff\ larger by about 100~K, practically at the same luminosity, since the inner 
structure does not change significantly with the EOS.\\ 
The results for different masses are shown in Fig.3, reporting the HR location 
of the models at 10Gyr. The differences are small for M$\ge$0.5\msun, where the 
regions of partial ionization do not dominate, and vanish at the lowest masses, 
because only the EOS+SCH is available for the physical conditions of their interiors.\\
Based on these results, we may conclude that use of both EOS leads 
to models with compatible effective temperatures and luminosities.
In Fig.\ref{figEOS3}, for the available masses, we report the location of the models 
of same chemistry computed by \citet{dotter2007}  with the Dartmouth Stellar Evolution 
Program (DSEP), using the FreeEOS by Irwin (2004) and otherwise a very similar input 
physics (L and \teff's are taken from their WEB page); for the same masses also the models 
by \cite{baraffe1997} are shown. Note that these latter models do not differ significantly from 
ours, both in L and \teff, while the models calculated with  DSEP at the lowest masses differ, 
especially in \teff, up to $\sim$300~K at 0.15M$_{\odot}$. This effect can be possibly attributed at the different EOS, but a detailed comparison of models would be required.

\subsection{Convection}
The ATON code allows us to model turbulent convection by adopting the 
traditional Mixing Lenght Theory (MLT, Bohm-Vitense, 1958) or the ``Full Spectrum of 
Turbulence'' (FST) model \citep{canuto1991,canuto1996} which takes into account the 
full eddies energy distribution (see Canuto e Mazzitelli, 1991 for a detailed 
description of the physical differences between the two models).\\ 
While for very low mass stars the description of convection
has no influence on the atmospheric structure, this does not hold for those masses in which
convection has a substantial degree of overadiabaticity, especially for the stars at
the turnoff of GCs. For these models a homogeneous modelling of convection in the 
atmosphere and the interior is highly recommended.
At \teff$>$4000K, the grids of models computed by \cite{heiter2002} by means of NEMO, 
a modified version of Kurucz's code, are available. These grids are provided both with MLT model and with the FST model by \cite{canuto1996}. 
A preliminary version of these latter grids has been used by \cite{montalban2001} and will be used 
in this paper (where possible).

\subsection{Atmospheric structure and boundary condition}

At \teff$\simlt$4000K, in the outermost layers of very low mass stars, radiative 
absorption is dominated by molecules, and the outcoming flux is far different  
from the frequency-averaged distribution provided by grey models \citep{baraffe1997,montalban2000}. 
At larger \teff, the atmospheric models become less critical, whereas the \teff\ itself
is heavily influenced by the treatment of overadiabatic convection. For the non-grey models
that employ an MLT treatment of convection in the atmosphere with a given $\alpha_{atm}$, not only the
$\alpha=l/H_p$\ used in the interior computation ($\alpha_{in}$) affects the \teff, but also
the optical depth at which the match between the atmospheric and the interior integration
is made, and the value of $\alpha_{atm}$ \citep{montalban2004}.
\cite{montalban2001} have shown that use of the NEMO grids of model atmospheres \citep{heiter2002}
computed with the FST convection may provide a good match to the interior models computed with
the same convection model in the interior, independently of the matching optical depth.

For the above two reasons we use, according to the stellar mass, boundary conditions based 
on two different grids of non grey models of atmosphere\footnote{The
same approach has been adopted for Pre Main Sequence stars  in \citet{dicriscienzo2008}}.
For M$\geq$ 0.5M$_{\odot}$ we use \citet{heiter2002} FST grids, and FST convection also
in the interior computations, whereas for M$\le$ 0.5M$_{\odot}$ we adopt the NextGen grids by 
\citet{hauschildt1999} computed by the PHOENIX code with 
the MLT treatment of convection and $\alpha_{atm}=l/H_p=1$. 
The available grids extend down to T$_{\rm eff}$ =800 K for the [M/H]=--2.0 models, but 
just down to 2000~K for larger metallicity. 
For these M$\le$ 0.5M$_{\odot}$ models, we adopt MLT convection also in the interior computation, 
setting $\alpha_{in}$=2.0\footnote{The choice of the $\alpha$\ parameter is less and less 
critical when decreasing the mass, as the external layers become so dense that 
convection becomes more and more adiabatic. Nevertheless, we use $\alpha_{in}$=2.0 to
allow for a smooth \teff\ transition between the upper (M$\geq$ 0.5M$_{\odot}$) and lower 
(M$\le$ 0.5M$_{\odot}$ ) MS models.}. 
Since the model atmospheres are computed assuming an ideal gas and a Saha--like thermodynamics, 
they should  not be used in the real gas domain where pressure effects are 
relevant, and a relatively small value of the optical depth is chosen for the match
between atmosphere and interior. We use $\tau_{ph}$=3(10) for M$<$($\geq$) 0.5M$_{\odot}$.
Notice that the grids of model atmospheres available are computed only for solar scaled mixtures. 
The lack of suitable model atmospheres for $\alpha$-enhanced populations  forces us 
to use for the boundary atmospheric conditions a grid with larger [Fe/H], to simulate
the $\alpha$--enhancement, following the procedure adopted by \citet{baraffe1997}. 
We use a grid for [Fe/H]=-1.7, obtained through interpolation between the  grids for 
[Fe/H]=-2.0 and -1.5, to compute $\alpha$--enhanced models with Z=0.0002.
\subsection{Transformations to observational plane}
In order to compare the models to the photometric data  in  NGC~6397 we convert 
luminosity, T$_{\rm eff}$ and surface gravity into absolute magnitudes and colors 
in the ACS filters. The method most used is to  calculate theoretical stellar 
spectra from atmosphere's models and convolving  these synthetic spectra with 
the filter transmission curves for a photometric system which defines the 
transmission of light through the  filter as a function of wavelength.  
Uncertainties derive especially from the missing or 
incorrect absorption features and simplifying physical laws, such the 
assumption of LTE. On the other hand, semiempirical  colors and bolometric corrections 
\citep[e.g.][]{vandenberg2003} have other uncertainties, e.g. they 
depend on the assumed distance \citep{dotter2007}.\\ 
For the specific case of ACS filters, we use the procedure by 
\citet{bedin2005}. They computed ACS color indices by using homogeneous set 
of ODFNEW model atmospheres and synthetic fluxes computed 
with Kurucz ATLAS9 code \citep{castelli2004}. 
Grids of magnitudes for different values of [Fe/H], for 3500K $\le$ 
T$_{\rm eff}$ $\le$ 50000K, 0 $\le$ log{\it g} $\le$ 5.0 
and microturbolent velocity 2.0 km/s$^{-1}$~ are provided. 
Visual bolometric correction BC$_{\rm V}$, visual magnitude M$_{\rm V}$, and 
color indices M$_{\rm V}$-M$_{\rm ACS}$ are given. The ACS 
magnitudes were computed by using the WFC/ACS transmission curves by 
\citet{sirianni2005}, while they adopted the V passband from \citet{bessel1990}. 
Finally, they assumed that Vega ACS magnitude would be equal to 0.00 in all 
passbands. \\ 
The bolometric corrections by \citet{bedin2005} for ACS filters do not extend below
T$_{\rm eff}$$=$3500K. For these low temperatures we can use the values 
obtained from the synthetic spectra of \citet{hauschildt1999}. Also in this case, 
the transmission filters of \citet{sirianni2005} were used, and zero points 
corrections to standard system are obtained from observed Vega spectra.\\
Obviously the availability of a unique set of colors-Teff\
transformations would be highly recommended; however, at least, for masses below
0.5 M$_{\odot}$, we have the advantage  of using the same bolometric corrections derived
from the atmospheric structures used as boundary conditions for the stellar models.\\
Finally we note that the  two sets of correlations  match very well in the main sequence, so that no
discontinuity arises in our transformed isochrones.
\begin{figure}
\centering
\vspace{-20pt}
\includegraphics[width=8cm]{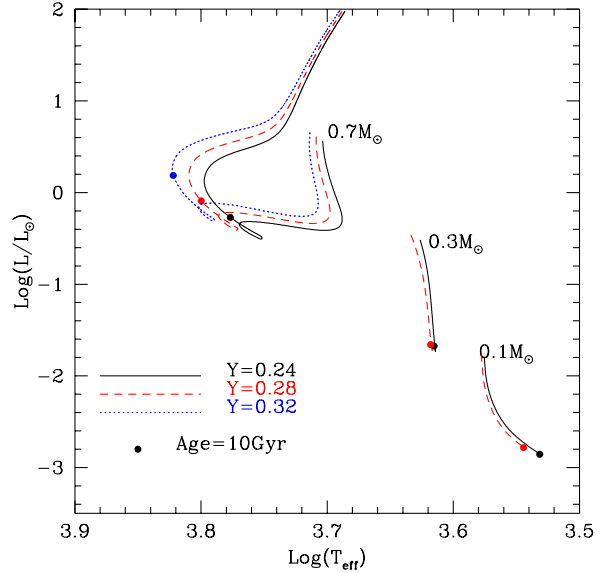} 
\vspace{-50pt}
\caption{Evolutionary tracks, starting from the pre-main-sequence, for three different masses with Z=0.0006 and 
different value of initial helium.The filled circle label the position in HR 
diagram of the star at Age=10Gyr. Only for the highest mass the tracks with the all  three Y are shown}
\label{figgraY}
\end{figure}

\section{Results: solar scaled models}
We computed evolutionary tracks for low mass objects (M$\le$0.8M$_{\odot}$) from the pre main 
sequence to the red giant branch, or until they reach an age of 20Gyr.  
Results are shown here for [M/H]=--1.5 and [M/H]=--2.0, but larger metallicities 
([M/H]=--1.00 and --0.50) are available upon request.\\ 
In this Section we present results for solar scaled mixtures \citep[][GS1999]{grevesse1999}, 
[$\alpha$/Fe]=0.4 models will be used in the Section 6 to compare with the data of NGC~6397.
Models are extended down to the HBMM when the atmospheric boundary conditions allow it.
 \\
We computed standard evolutionary tracks with an initial helium content close 
to the Big Bang abundance \citep[that is Y=0.24,][]{coc2004}, and 
models with larger helium (Y=0.28, 0.32 and 0.40). From the evolutionary tracks, isochrones
are derived for typical ages expected for GCs, from 10 to 14 Gyr.\\
Fig. \ref{figgraY} compares 
the HR diagram evolution for three different masses (M=0.70, 0.30 and 0.10 M$_{\odot}$) 
and different Y.  Models with a larger helium abundance  
have larger luminosity and \teff, due to the average larger mean molecular weight
$\mu$. This effect is more evident in the stars with a radiative core (here shown is the M=0.7\msun) 
than in the totally convective stars like the 0.3\msun. The difference 
increases again in the lowest masses (\simlt 0.15 M$_{\odot}$), where 
partial degeneracy begins playing a role.

\subsection{Mass-luminosity relation}

\begin{figure}
\centering
\vspace{-10pt}
\includegraphics[width=8.5cm]{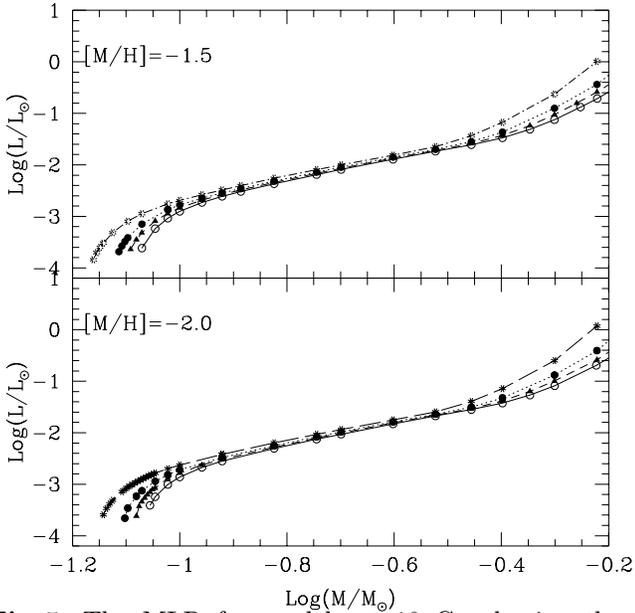} 
\vspace{-10pt}
\caption{The MLR for models at 10 Gyr  having the labelled value of metallicity 
and different helium abundance (=0.24-open circles, =0.28-filled triangles, =0.32-filled circles, =0.40-asterisks). In each case the lower masses are the derived HBMM and reported in Table 1.}
\label{figmlr}
\end{figure}

\begin{figure}
\centering
\includegraphics[width=8.5cm]{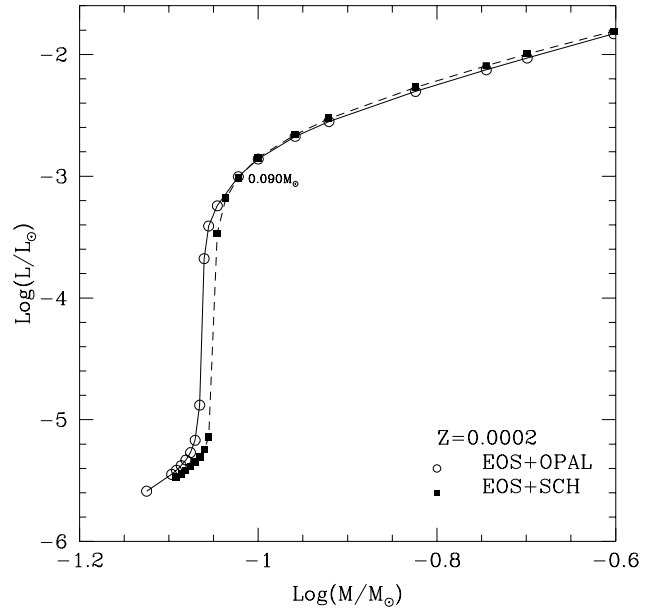} 
\caption{Comparison at very low masses between the MLR obtained using models computed 
with different equations of state (see text).}
\label{figmlrEOS}
\end{figure}
\begin{figure}
\centering
\vspace{-20pt}
\includegraphics[angle=0,scale=.4]{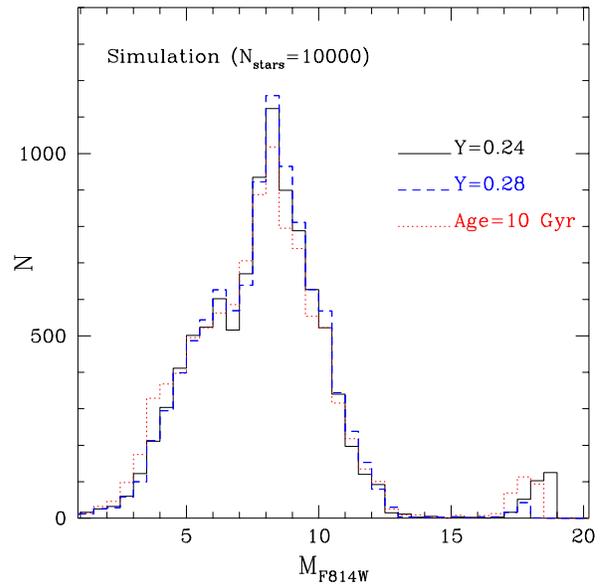}  
\vspace{-50pt}
\caption{Theoretically predicted luminosity function for the models at Age=13 Gyr and 
with Z=0.0002 and two different helium abundance. A power law with exponent  
$\alpha$=--0.5 was used as a mass function. Also the LF at Age=10 and with Y=0.24 is shown for comparison. }
\label{figlfteo}
\end{figure}

The mass-luminosity relation (MLR) is essential for the 
comparison with the data, as it enters in the conversion 
of the (assumed) mass function into the luminosity function (LF). \\ 
Any change of slope of the MLR will be reflected in the luminosity function,
defined as dN/dL=dN/dM $\times$ dM/dL.  
Where the MLR presents an inflection point, the LF has a relative maximum or minimum.
As the luminosity decreases along the MS, there are two inflection points, 
responsible for two main peaks in the LF (see Fig.\ref{figlfteo}): the first, 
at M$_{\rm F814W}$ $\sim$ 6 mag, is due to the transition between pure MS models and
models that suffer the effects of evolution (and are more luminous in the MS due to 
the hydrogen consumption). Therefore, the corresponding peak in the LF is a function 
of age, and, we will see, of Y. 
After a small range of homologous models,
the MLR steepens progressively due to the onset of molecular absorption  in the 
stellar envelope; this produces a
gradual increase in the LF. When models become fully convective, at $\sim$0.35M$_{\odot}$,
the MLR relation begins to flatten again, and the presence of this
inflection point results in the large peak at M$_{\rm F814W}$ $\sim$ 8 mag,  
present in all the GC LFs \citep{dantona1998} and shown in Fig. \ref{figlfteo}.\\
Finally, at masses M $\sim$0.12M$_{\odot}$, the MLR relation flattens even more. This is
due to the onset of degeneracy in the core of the star \citep{dantona1998}, and 
this latter decrease in the LF is dependent on the EOS.
The HR diagram does not show the minute features of the MLR derivative, but it shows two ``kinks",
the first one corresponding to the onset of molecular hydrogen dissociation in the envelope
\citep{copeland1970} and the second to the onset of degeneracy.\\
In Fig. \ref{figmlr} we show the dependence of the MLR on the metallicity 
and helium content at 10 Gyr. Y influences mostly the evolved part of MS and the location 
of the very low masses, where degeneracy sets in and the second MS kink is located. 
At the MS end, decreasing the mass, the higher helium models remain more luminous and 
hotter. These features will affect the low luminosity LF, that will decrease 
more slowly with decreasing luminosity for larger Y.
At the largest MS luminosities, the larger is Y, the smaller is the slope of the MLR 
of those models that partially burn their hydrogen during a Hubble time, 
with respect to the case of Y=0.24. 
This is an obvious feature of the evolutionary models: the larger Y produces larger
MS luminosities and faster MS evolution.\\
The quantitative results concerning the lowest luminosities will also depend on the EOS, as
we can easily understand by comparing the MLR relations obtained using EOS+OPAL 
and EOS+SCH tables (Fig. \ref{figmlrEOS}).
For masses larger than 0.1M$_{\odot}$ the models using EOS+SCH are slightly more luminous, 
as a consequence of the differences in \gad, but the trend is reversed closer to degeneracy. 
Obviously this variation in the slope of the MLR will produce a different shape in the peak of the LF,
and in particular we expect a stronger peak when EOS+SCH are used. \\
An interesting theoretical feature of the models at the boundary between
low mass stars and brown dwarfs is shown in Fig. \ref{figmlrEOS}. This
figure refers to the Z=0.0002 case, for which an extended atmospheric grid is available.
\begin{table}
\begin{center}
\caption{Hydrogen burning minimum mass (HBMM) for Z=0.0002 and  Z=0.0006, [$\alpha$/Fe]=0, and  different Y.}
\begin{tabular}{|c|c|c|c|c|}
\hline
&Y=0.24&Y=0.28&Y=0.32&Y=0.40\\
\hline
Z=0.0002&0.088M$_{\odot}$&0.083M$_{\odot}$&0.079M$_{\odot}$&0.072M$_{\odot}$\\
Z=0.0006&0.085M$_{\odot}$&0.081M$_{\odot}$&0.077M$_{\odot}$&0.069M$_{\odot}$\\
\hline
\end{tabular}
\end{center}
\end{table}
In between the HBMM ---the smallest mass that stabilizes in MS, in a configuration 
in which the total luminosity is provided by the proton--proton (p--p) reactions in the core---
and the pure brown dwarfs ---that never ignite the p--p chain--- there is a small range of masses
for which nuclear burning contributes to the stellar luminosity for even several billion
years, but in the end these objects finally cool as brown dwarfs. This is a common occurrence 
in population I \citep[the transition masses defined in][]{dm1985}, 
where the MS merges without discontinuities into the brown dwarf
cooling sequences. On the contrary, the MS of the population II has a much sharper drop, because the
much smaller opacities put the HBMM at minimum luminosity a factor $\sim$10 larger, so that the
transition masses cover a very small mass range. There is then a ``luminosity gap" between the
end of the MS and the luminosity at which the smaller brown dwarfs are able to slow their cooling
down to the typical age of population II stars (10--12 Gyr).
Fig. \ref{figmlrEOS} translates into possible LF. In   Fig.\ref{figlfteo} LFs for 10 and 13 Gyr are 
plotted, assuming a power law MF with exponent $\alpha$=--0.5.
At 13 Gyr, the low mass brown dwarfs should emerge as a small peak at M$_{\rm F814W}$ $\sim$18, corresponding 
to near infrared magnitudes $\sim$30 at the distance of NGC~6397. The peak is 
$\sim$1 mag brighter if the age is 3 Gyr smaller. 
We regard this prediction as an educated guess on the possibility that the dimmest luminosities regime 
is populated not only by white dwarfs (Richer et al. 2008) but also by very cool brown dwarfs.
However, dynamical models  indicate that these objects should be preferentially stripped from the cluster. Since there is a strong difference in these two populations, as the cool white dwarfs would be located at a color 
M$_{\rm F606W}$-M$_{\rm F814W}$ $\sim$1-1.2mag, while the cool brown dwarfs will not be visible in the
F606W band, having M$_{\rm F606W}$-M$_{\rm F814W}$ $\sim$5.5-7mag only future observations with even more capable telescopes should clarify the question.

\section{$\alpha$-enhanced models, FG and SG populations, CNO enrichment}
\begin{table*}
\begin{center}
\caption{Descriptions of the chemistry  of the models calculated for this work.}
\begin{tabular}{|c|c|c|c|c|c|c|}
\hline
Name & Description & Z & [C/Fe] & [N/Fe] & [O/Fe] & Reference\\
\hline
\hline
\multicolumn{7}{|c|}{Z=0.0006} \\
\hline
CNOx1s & Solar-scaled      &0.0006&0.00 &0.00 &0.00   &GS1999\\
CNOx1a &[$\alpha$/Fe]=0.4  &0.0006&0.00 &0.00 &0.40 &GS1999\\
\hline
\hline
\multicolumn{7}{|c|}{Z=0.0002} \\
\hline
CNOx1s                   &Solar-scaled&0.0002&0.00 &0.00 &0.00   &GS1999\\
CNOx1a                   &[$\alpha$/Fe]=0.40 &0.0002&0.00 &0.00 &0.40 &GS1999\\
CNO$\uparrow$            & total(CNO)=1.6      &0.0003&0.00&1.40&0.20&this work\\
\hline
\end{tabular}
\end{center}
\end{table*}

\begin{figure}
\centering
\includegraphics[width=6.5cm]{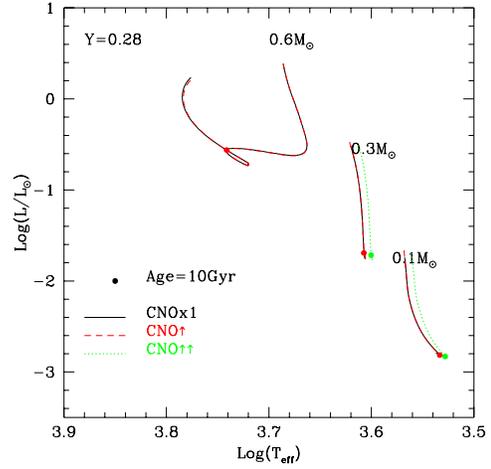}
\caption{ We show the comparison between the diagram to the standard 
CNO tracks at different masses(full line) and the ``peculiar'' ones calculated (dashed and dotted line) with the evolution of the same mass.}
\label{diffCNO}
\end{figure}
\begin{figure*}
\centering
\vspace{-250pt}
\includegraphics[width=18cm]{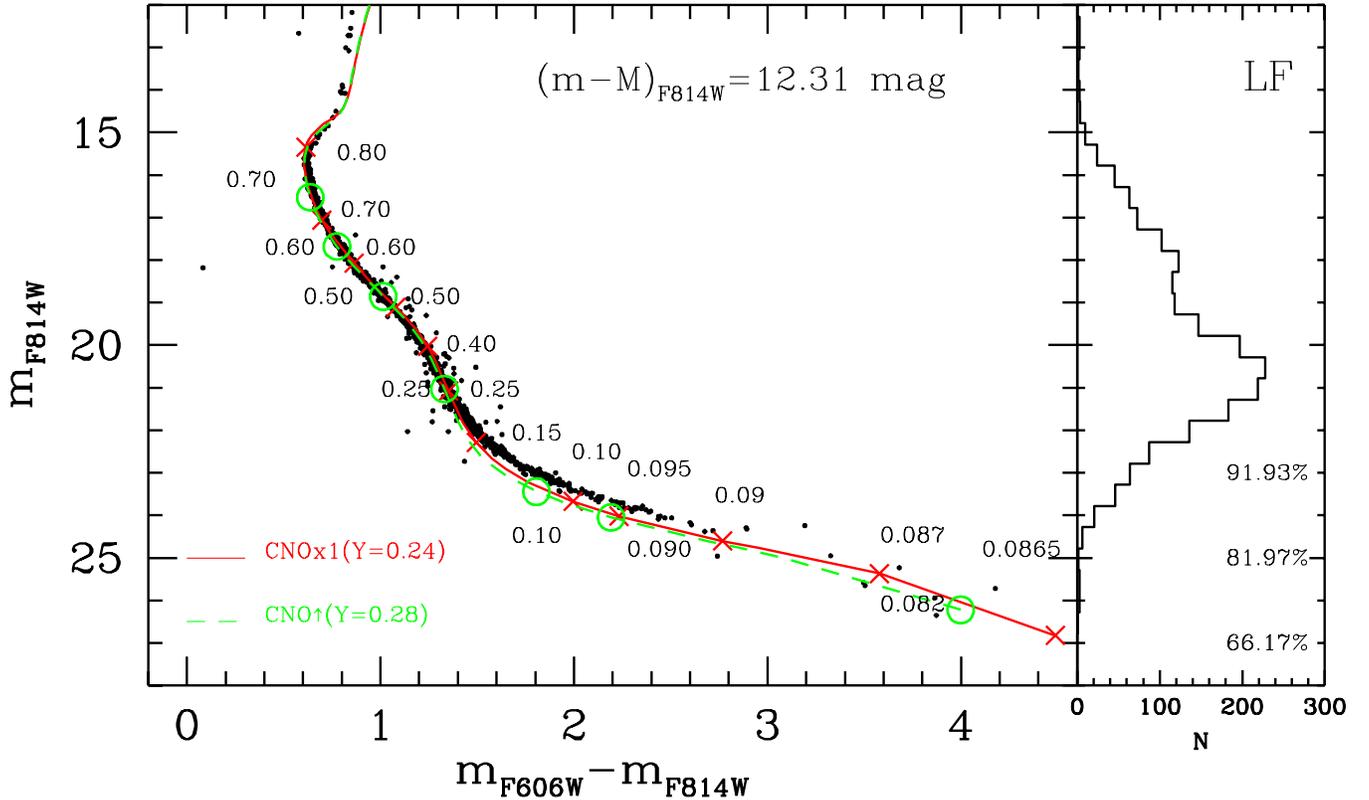}
\vspace{-120pt}
\caption{{\it Left panel} Main Sequence of NGC~6397 from 
\citet{richer2008}. Also shown are the best fit isochrones for FG (red) and SG (green) populations. Numbers indicate stellar masses (up-FG,down-SG) of same 
models (crosses-FG, open circle-SG) in M$_{\odot}$ units. {\it Right panel}: NGC~6397 MS Luminosity 
Function from data; for three low luminosities are reported the completeness  
fraction taken from Table 4 of \citet{richer2008}.}
\label{figCMD}
\end{figure*}

Comparison with low metallicity cluster stars requires use of models with $\alpha$-enhanced  mixtures,
for which we adopt [$\alpha$/Fe]=0.4. 
As  discussed only up to $\sim$30$\%$  of stars of NGC~6397 can belong to the FG, 
the majority show indeed the Na--O anticorrelation, and many stars have very high Nitrogen content.
Therefore we need also models to represent the SG. We will assume that it may differ both 
in helium content and in total CNO content from the standard FG. 
A resonable assumption for CNO is that we assume an overbundace of  N by about 1.4 dex, 
and a variation of --0.2 dex for O, leaving carbon unchanged (A. Bragaglia, private comunication). 
The total ``metallicity'' Z in mass fraction for this CNO-enhanced mixture 
(CNO$\uparrow$) is now Z=0.0003. 
We used the OPAL Web tool to compute on purpose radiative opacities for this mixture. For T$\le$15000K we still use the opacities by \citet{ferguson2005}  as lower temperature opacities do not affect the structure of the models we are considering.
We compute on purpose radiative opacities for this mixture.
All models are computed for helium mass fraction Y=0.24 and Y=0.28 and summarized in Table 2.\\
In general, as discussed in the analysis by \citet{ventura1851}, the largest differences 
between standard and CNO--enhanced mixtures  are found in  the ionization zone  
of the CNO elements, but in this case the differences are very small since the variation 
is not very large, and the initial abundances are very small. 
As a result, pratically no differences are found in  effective temperature and luminosity of the models (see Fig.\ref{diffCNO}). \\
Another possible hypothesis is that Oxygen in the SG is basically not very different from
the FG value, as most of the \cite{carretta2009a} measurements for O are only upper limits.  
In this case, the CNO abundance becomes even larger. In this case, we adopt as boundary conditions the
grid [M/H]=--1.5. (CNO$\uparrow$$\uparrow$ models, see Table 2). As reported in Fig.\ref{diffCNO} also in this case the differences are very small but for masses of about 0.3M$_{\odot}$ the temperature are little smaller.

\section{Comparison with the data of NGC~6397}
We analyse both the CMD diagram and the luminosity function simulations. 
\subsection{Color-Magnitude diagram}
In Fig. \ref{figCMD} we compare the models with the  deep photometry of NGC~6397 by \citet{richer2008}. 
We plotted the isochrone that allows the best fit of both the low main sequence and the TO 
for the labelled value of distance modulus and reddening, with [Fe/H]=-1.99 dex \citep{carretta2009c}  and [$\alpha$/Fe]=0.4 (solid line--red in the electronic version) and with 
an age of 12 Gyr which is comparable with the age obtained from the White Dwarf Cooling Sequence of NGC 6397 by \citet{hansen2007}. The distance modulus in F814W photometric  band  is compatible with the true distance modulus (=12.03 mag) and  E(F606W-F814W)(=0.20 mag) reported in literature \citep[see for example Table 3 of][]{richer2008}.\\ 
As already found by \citet{richer2008},  below 0.20 M$_{\odot}$ the isochrone and the data 
do not match perfectly, however the  data appear to terminate at about the magnitude predicted by models.
We then confirm the suggestion of \citet{richer2008} that they have observed the termination of 
the hydrogen burning sequence. \\
In Fig.  \ref{figCMD} we also show (dashed line-green in the electronic version) the position of the isochrone obtained with models computed with Y=0.28 and CNO$\uparrow$ using the same distance modulus and age used for the FG isochrone. {The two isochrones are very similar except around m$_{\rm F814W}$=23 mag where they deviate and  the FG isochrone is a little brighter. This aspect is important to understand if the  tightness of the MS at this interval of magnitude depends on observational error only or is the consequence of the presence of a second generation made of stars with higher helium abundance and CNO enhancement.
We select data within three different intervals of half magnitude below the MS turnoff, and rectify their colors by
subtracting the color of their best-fit line, as shown in Fig. \ref{fighel}, left and medium panels. 
We do not consider the possible presence of binaries, as \citet{davis2008} have shown 
that NGC~6397 has a primordial binary fraction of only $\sim$1$\%$.
The histogram of the color displacements from the best fit line, shown in the right
panels, are fitted with a gaussian profile with the labelled $\sigma_{obs}$.\\ 
In the same panel, for each interval of magnitude we report the color displacements for synthetic populations (assuming the  distance modulus for NGC 6397 reported in Fig.\ref{figCMD} ) under the hypothesis 
that the 30 $\%$ of all stars are primordial (standard  helium and CNO abundance), and the restant 70 $\%$ is composed by a mixture of SG population with CNO$\uparrow$ and helium respectively up to Y=YUP=0.25, 0.26 and 0.28.
As expected, the largest difference between $\sigma$ is obtained in the third interval of magnitude. From the values of dispersion reported we obtain that a SG made of stars with an helium dispersion of $\Delta$Y=0.02 is compatible with the tightness of the MS of NGC~6397.\\
In the case of  CNO$\uparrow$$\uparrow$ models (see Section 5) the isochrone calculated for  Y=0.28   overlaps exactly on the FG's one. In this case the observational spread of MS at low magnitude may suggest an even larger spread of helium between FG and SG. 
\begin{figure}
\vspace{-120pt}
\centering
\includegraphics[width=9.5cm]{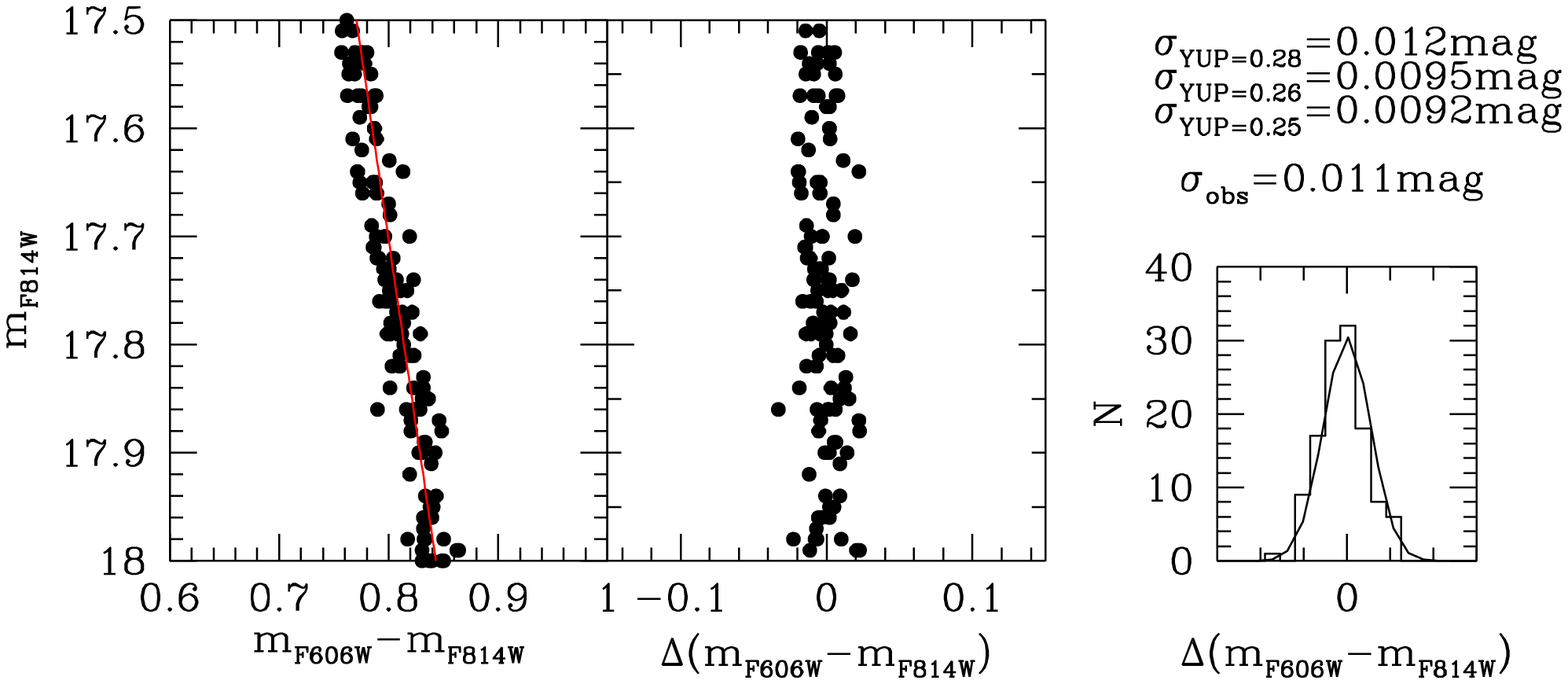}
\vspace{-150pt}
\vspace{-80pt}
\centering
\includegraphics[width=9.5cm]{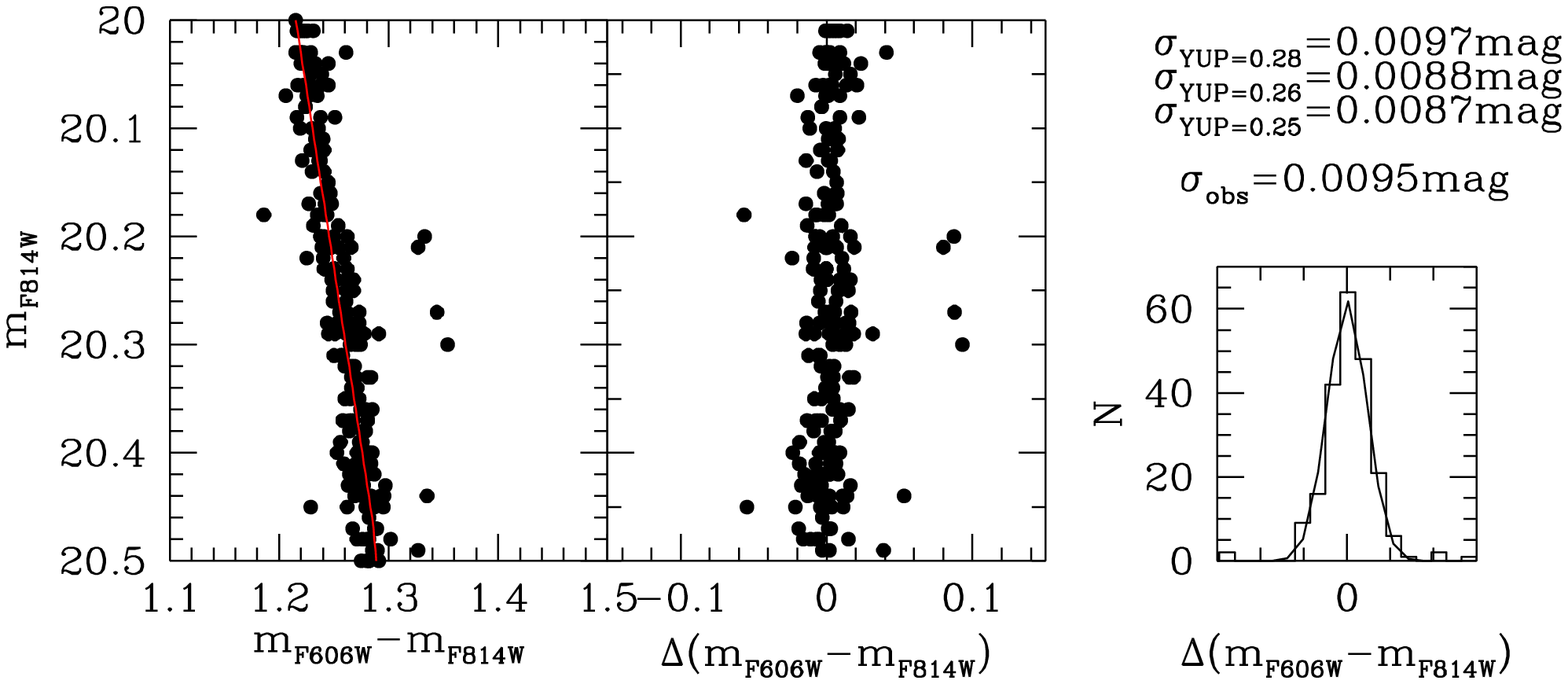}
\vspace{-150pt}
\vspace{-80pt}
\centering
\includegraphics[width=9.5cm]{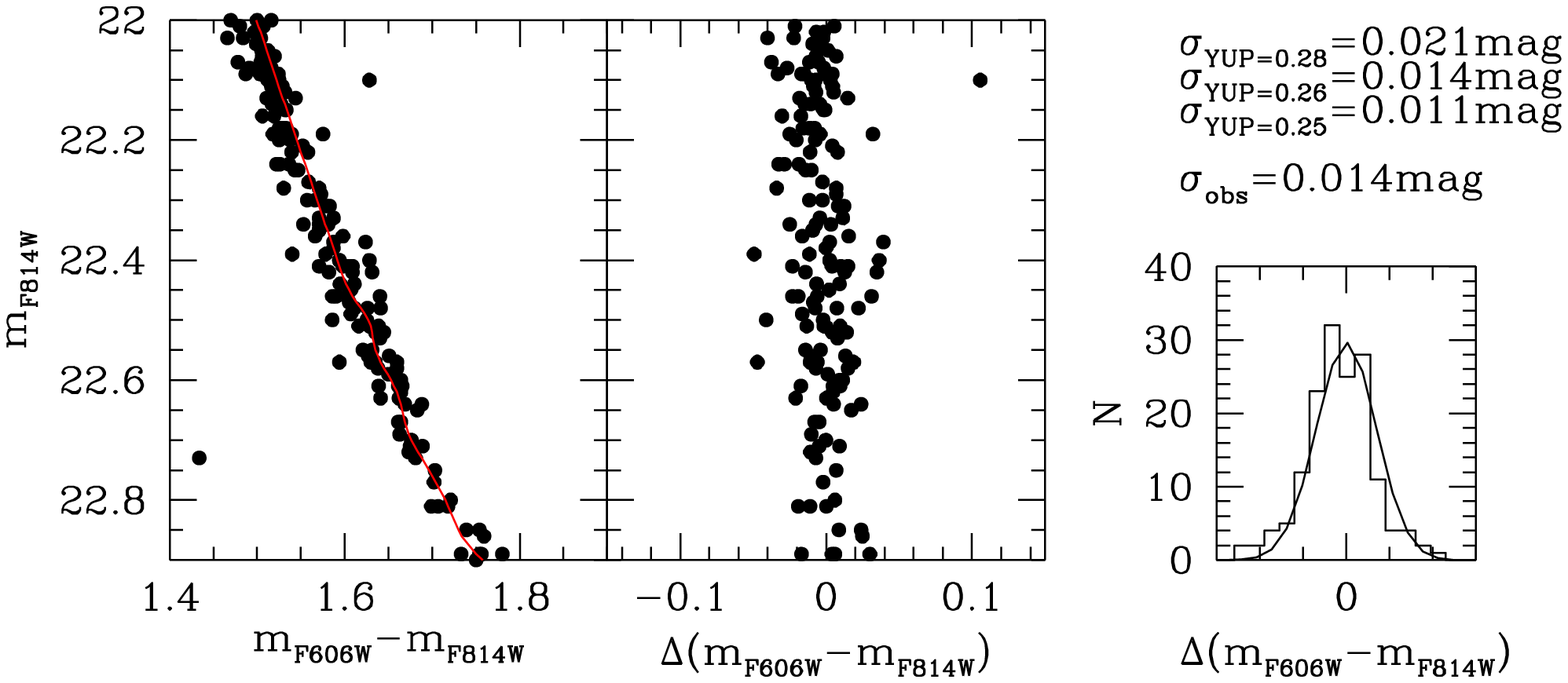}
\vspace{-80pt}
\caption{{\it a) } In each row a little portion of the MS of NGC 6397 is shown
on the leftmost side, while in the middle panel the magnitude m$_{F814W}$ is reported
versus the color distance from the best--fit line. On the right panel we show the histogram of the color 
displacements, fitted with the best fit gaussian curve for the data. We also report the $\sigma$ of the best fit gaussian curve obtained for  of synthetic population made of 30 $\%$ of stars belonging to FG and the restant 70 $\%$ made of stars of SG with different YUP.}
\label{fighel}
\end{figure}

. 

\subsection{Luminosity function}
We now compare the observed luminosity function to the theoretical simulations. We use the magnitudes in  
F814W filter and derive synthetic populations from  models at different ages, in the
plausible range from 9 to 14 Gyr.\\ 
The mass-luminosity relation from our models was discussed in Section 4.1; now we discuss the 
choice of the mass function (MF). At the beginning we consider this cluster as formed by an unique population in order to define a method  to compare theoretical and observational luminosity function. Due to the dynamical evolution of the cluster, the present MF is not the initial one. 
NGC~6397 has a collapsed core \citep{DK1986}, 
as  it has evolved past the potentially catastrophic phase of core collapse, 
and is dynamically old. In particular it  has been shown that NGC~6397  exhibits 
mass segregation, which certainly has affected the  MF \citep{hurley2008}.
\citet{silvestri1998} have shown, adopting their own 
low mass models and those by \citet{baraffe1997}, that the bulk of the MF can 
be described by a unique power law of the form dN/dM=k M$^{\alpha}$ with index 
${\alpha}$=--0.5. \citet{richer2008} found that  ${\alpha}$=--0.13 gives  
the highest $\chi^2$ when comparing the models with their data of NGC~6397. 
They found  an even better result when a lognormal distribution is used,  which have the advantage to truncate the LF at the extreme low mass end but the disadvantage to  introduce a new parameter.  
\begin{figure}
\centering
\vspace{00pt}
\includegraphics[angle=0,scale=.30]{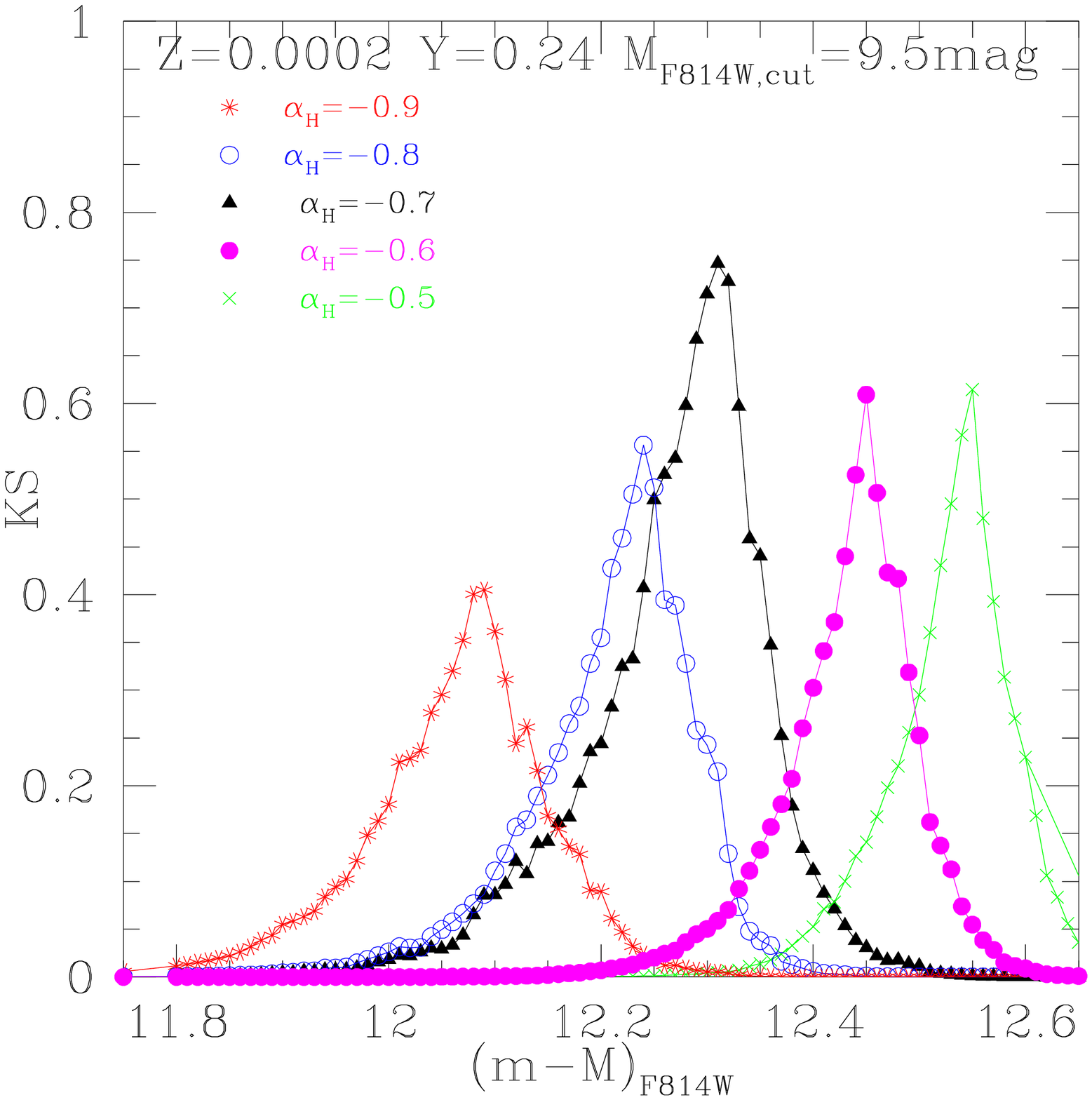}
\vspace{-30pt}
\includegraphics[angle=0,scale=.30]{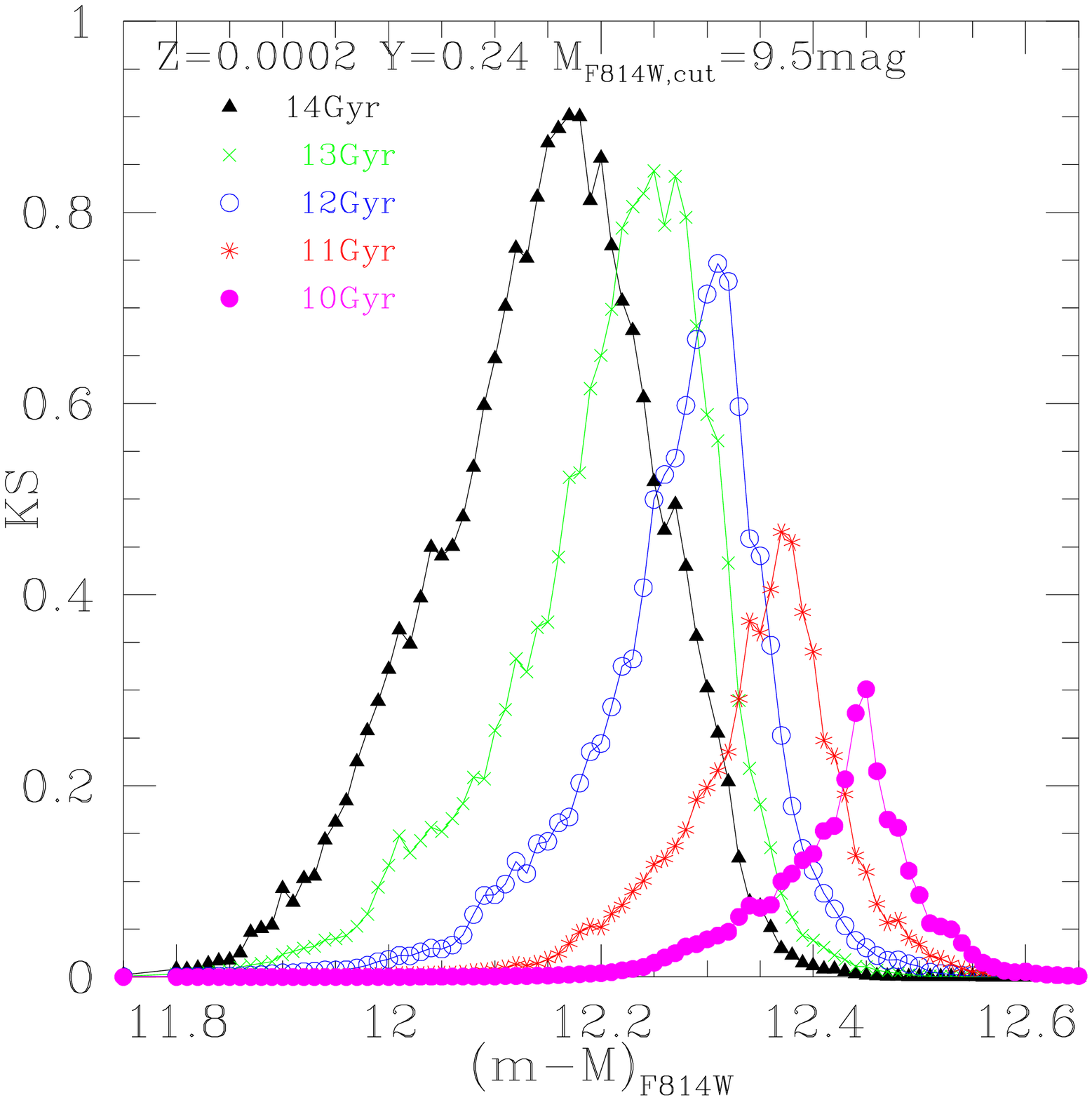}
\includegraphics[angle=0,scale=.30]{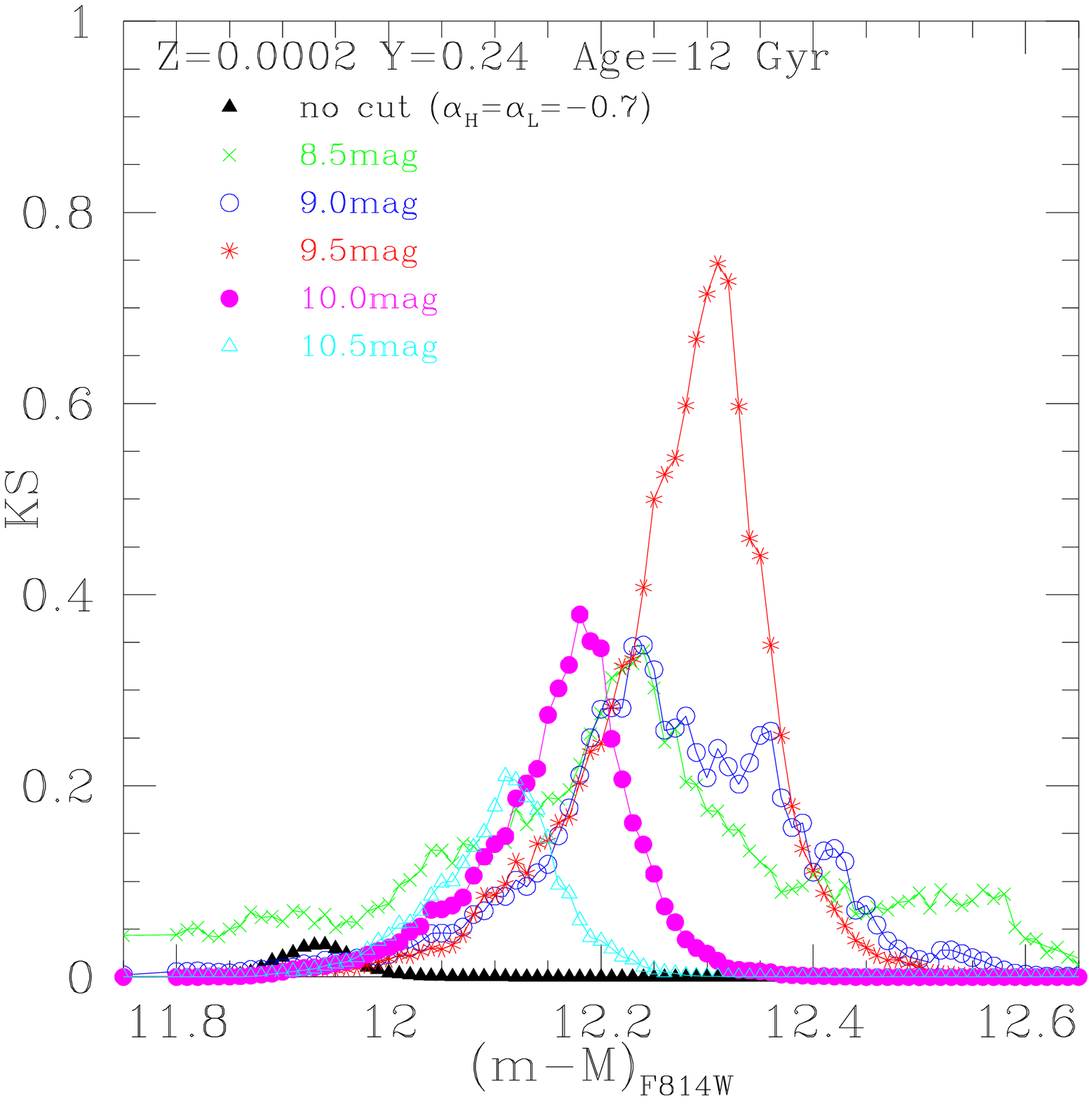}
\vspace{-40pt}
\caption{Distribution of KS with distance modulus for the labelled value of Z and 
Y and for different $\alpha_H$ (({\it upper-panel})), different ages ({\it medium-panel}) and M$_{\rm F814W, cut}$ ({\it down-panel})
(see text). In all cases a double power law was used with $\alpha_L$=--0.1. In the medium and down panels $\alpha_H$=--0.7 as inferred by the upper panel is used.  }
\label{figKS}
\end{figure}
A similar result was obtained by \citet{demarchi2000}which  suggested that  no single power law distribution  is compatible with the 
MS of NGC~6397; in particular they  found  that the MF is less steep for M$\le$0.3M$_{\odot}$.\\ 
We simulate the synthetic populations using a power law MF with two different slopes above ($\alpha_H$) and below ($\alpha_L$) a cutoff 
magnitude in the range M$_{\rm F814W, \rm cut}$=8.5-10.5 mag (here we consider a single MF as a particular case with $\alpha_H$=$\alpha_L$.); this cutoff magnitude, for the
chosen distance modulus, corresponds to a ``cutoff mass''.  
At a fixed age, MLR gives the magnitudes of each extracted mass, according to the chosen MF, which depends by four parameter ($\alpha_H$, $\alpha_L$, M$_{\rm F814W, \rm cut}$ and distance ). 
We simulate photometric errors  considering gaussian errors for the magnitudes of each extracted mass in order to reproduce the width of the upper main sequence. In addition we have taken into consideration the uncertainties in completeness, by multiplying the random extractions in a given interval of masses for the completeness fractions given in Table 4 of \citet{richer2008} and determined with the artificial star test described in \citet{anderson2008}.\\
We have then used the Kolmogorov-Smirnov test to compare the  observed luminosity function of NGC~6397 with  the theoretical one, which depends on the MF used to extract masses ($\alpha_H$,$\alpha_L$M$_{\rm F814W, \rm cut}$), on age and distance modulus.
This statistical method has the advantage  of being non-parametric and without making  assumptions about the distribution function of the data, it returns the probability   that two arrays of data values are drawn from the same distribution. The scalar KS\footnote{We use  the {\tt kstwo}  algorithm from "Numerical Recipes" ,3nd edition.} (varying  between 0 and 1) which give  the significance level of the KS statistic  i.e. the probability with which  we can accept the null hypothesis.
KS=1 means that  the simulated and observed data follow  the same function.\\
Another great advantage of this statistical method is that the two arrays of data does not need to have the same number of elements; this means that we can build our synthetic populations with a greater number of stars then the real stars from which we made the observed LF, making the results independent from the  random extraction.
We have studied the dependence of KS numbers as a function of distance modulus  for each of the four free parameter (age, $\alpha_H$,$\alpha_L$ and M$_{\rm F814W, \rm cut}$) used to build the synthetic population and then  we have compared the results in order to choice the best fit parameter. In particular as done by Richer et al. (2008) for their $\chi^2$ method both  $\alpha_H$ and $\alpha_L$ were allowed to range from -1 and 1 in 200 steps considering. We have explored the case  of a  single MF as a particular case of this situation ($\alpha_H$=$\alpha_L$). We find that the best combination of power law exponents  for, respectively, higher and lower masses, are $\alpha_H$=--0.7 and $\alpha_L$=--0.1. The upper panel of Fig. 11, where for example are shown the variation of  KS with distance modulus for different selected  $\alpha_H$, justify our choice. We also note that the distance modulus for which we have the higher probability is about   the same  distance modulus  for which we have the best comparison between isochrones and CMD
(see Fig.\ref{figCMD}), confirming the validity of our method.
The same result is obtained using models with higher helium abundance.
For these values, in Fig. \ref{figKS} we also report KS as a function of  distance modulus 
for different ages (medium  panel) and for different cutoff  magnitudes (lower panel).
In the bottom panel, also the case of a single power law with index --0.7 is shown (black triangles): 
we see that  a unique power law does not give a good match to the observed LF. 
Notice however that for higher masses this value is very different from the one obtained by \citet{richer2008} (=--0.13).
However we want too stress  that this our result is consistent with their consideration  that a lognormal function produces a better ${\chi}^2$  values than the best fitting single power law MF. In fact when a lognormal function is used one has two adjustable parameters as in our case.
Concerning the age, as shown in medium panel of Fig. 11, the best agreement with observations is obtained for an age of 14-13 Gyr, but if we also consider the best distance moduli and take into account the CMD  we can choose 12 Gyr together with  M$_{\rm F814W, \rm cut}$=9.5 mag,  
corresponding to $M$$\sim$0.18M$_{\odot}$\footnote{This value is much lower 
then the mass ($\sim$0.3M$_{\odot}$) found by \citet{demarchi2000} }.\\
In Fig. \ref{figDISTRI} we report the NGC~6397 MS LF compared with the best 
fitting double power law mass functions. The overall agreement is satisfactory.
Only at M$_{\rm F814}$$\ge$12mag ($\sim$ M=0.090M$_{\odot}$), the observed stars 
are fewer then predicted; this may mean that the MF is even flatter at these lowest masses, due to more effective evaporation, but it may also be due to some deficiency in the models.
\begin{figure}
\centering
\vspace{-20pt}
\includegraphics[angle=0,scale=.4]{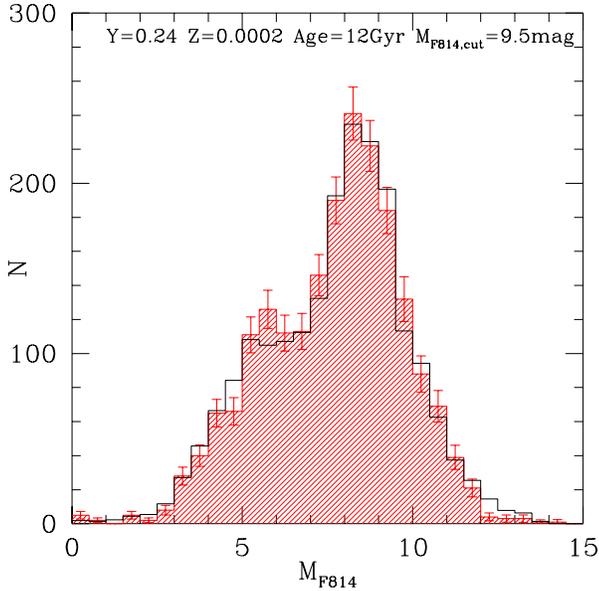}
\vspace{-50pt}
\caption{Comparison between observed and calculated distribution  of MS stars of NGC~6397 for the labelled values of the parameters. The power law used is the same of Fig. \ref{figKS}. The error bars take into account the Poisson's error  and incompleteness corrections.}
\label{figDISTRI}
\end{figure}
\begin{figure}
\centering
\vspace{-20pt}
\includegraphics[angle=0,scale=.4]{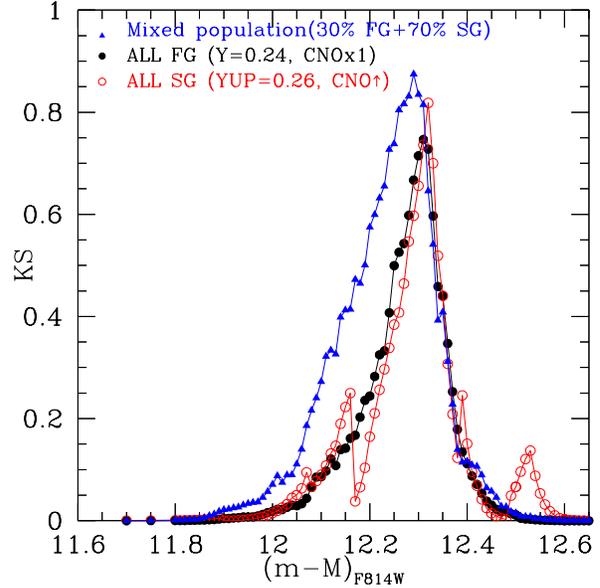}
\vspace{-50pt}
\caption{Distribution of KS with distance modulus for different synthetic populations.}
\label{KSY24vs28}
\end{figure}
In Fig. \ref{KSY24vs28} we also show the distribution of KS in the case that 30$\%$ of stars belong to FG (Y=0.24 and CNOx1a) and 70$\%$ to the SG (YUP=0.26 and CNO$\uparrow$) which give the best match of the widht of the MS (see Fig. \ref{fighel}) and in the case  that all stars belong to SG. The conclusion is that, since comparable KS, and reliable distance moduli are obtained 
in all cases, as from the widht of MS also from the luminosity function  we cannot exclude that the cluster contains either a
mixture of stars with different helium, or a single helium abundance  for all stars.

\section{Conclusions}
We have computed new models  for the main sequence down to the hydrogen burning minimum mass,
adopting two different version of an updated equation of state   and made simulations of the luminosity
functions for different choices of the mass function and the initial helium content.
The results are compared with  the recent observations of the MS of  NGC 6397
 by Richer et al. (2008). Using a Kolmogorov Smirnov test to compare observed and simulated LF we
found that a  double power law  for the mass function well reproduces the observed luminosity function 
in F814W photometric band.
However, both the  models for simple or mixed population according the spectroscopic data provide a good fit of LF. A stronger results is obtained from the analysis of the width of MS from which we find that in any case any  helium variations must be confined within $\Delta$Y=0.02 in the case of CNO overabundance predicted by a mixing between 50$\%$ of pristine gas and 50$\%$ of gas eject by 5\msun AGB stars as suggested by Ventura $\&$ D'Antona (2009). Instead we find that a arger spread ( 0.02$\le$$\Delta$Y$\le$0.04) in helium between primordial and intermediate generation is compatible with the widht of main sequence when CNO$\uparrow$$\uparrow$ models are considered.
\\

The complete sets of isochrones transformed for ACS filters F814W and F606W, 
calculated for this work, are available upon request to the authors 
and will be soon inserted on WEB at site http://www.mporzio.astro.it/$\%$7Etsa/

\section*{Acknowledgments}
We thank S. Cassisi for providing the  color-T$_{\rm eff}$ 
transformations and J. Anderson, A. Bragaglia, A. Dotter, A. Milone and H. Richer and G. De Marchi for useful discussions. We also thank the referee for the extensive and critical review of our paper.\\
Financial support for this study was provided by MIUR under the PRIN 
project ``Asteroseismology: a necessary tool for the advancement in the study 
of stellar structure, dynamics and evolution'', P.I. L. Patern\'o 
and by the PRIN MIUR 2007 
``Multiple stellar populations in globular clusters: census, characterization and
origin".
 
{}
\end{document}